\documentclass
[twocolumn,prl,epsfig,showpacs,preprintnumbers,amsmath,amssymb,showkeys]{revtex4-1}
\usepackage{graphicx}

\begin{document} \bibliographystyle{apsrev}
\title{Entanglement Entropy and Mutual Information Production Rates \\ 
in Acoustic Black Holes} \author{Stefano Giovanazzi} 
\email{Electronic address: stevbolz@yahoo.it} 
\affiliation{Kirchhoff-Institut f{\"u}r Physik, Universit{\"a}t Heidelberg, 
Im Neuenheimer Feld 227, 69120 Heidelberg, Germany} \date{\today}

\begin{abstract}
A method to investigate acoustic Hawking radiation is proposed, where entanglement 
entropy and mutual information are measured from the fluctuations of the number of 
particles. The rate of entropy radiated per one-dimensional (1D) channel is given by 
$\dot{S}=\kappa/12$, where $\kappa$ is the sound acceleration on the sonic horizon.
This entropy production is accompanied by a corresponding formation of 
mutual information to ensure the overall  conservation of information.
The predictions are confirmed using an \emph{ab initio} analytical approach 
in transonic flows of 1D degenerate ideal Fermi fluids.
\end{abstract}
 \maketitle

Hawking's prediction of black hole evaporation \cite{Hawking74} is regarded as 
one of the most remarkable discoveries of modern theoretical physics. 
Black holes radiate like black-bodies with a temperature proportional to their 
surface gravity $\kappa$. As a consequence of Hawking radiation, 
black holes lose mass and eventually disappear. 
Hawking's prediction confirms Bekenstein's conjecture that black holes possess 
intrinsic entropy \cite{Bekenstein73}, which would be exhausted by radiation during the 
evaporation process. Black hole entropy has inspired the development of new concepts, 
such as entanglement entropy \cite{Bombelli86,Holzhey94}, which is today of increasing 
relevance for quantum information and renormalization group techniques \cite{Klich09}.

Hawking radiation itself does not require gravity to exist. It occurs 
in various analogue models of black holes \cite{GravitationalAnalogues}, 
in particular, in acoustic black holes, the first suggested model systems consisting 
ideally of irrotational Euler fluids turning supersonic \cite{Unruh81}.
These are also called \emph{dumb holes} as sound generated inside 
the supersonic region (the analogue of the black hole interior) does not propagate 
back into the subsonic region. Based on the correspondence between the equations 
of motion for sound near his sonic horizon and those of a quantum field propagating 
near the event horizon of a black hole, Unruh predicted \cite{Unruh81} the emission 
of thermal sound from the sonic horizon.
The temperature of the radiated 
sound is given by Hawking's celebrated formula \cite{Hawking74}
\begin{eqnarray}
  k T_{\text{bh}} = \frac {\hbar  \kappa}{2 \pi}.
\label{gen_Hawking}
\end{eqnarray} 
Hereinafter, the Boltzmann constant $k$ is attached to any temperature symbol, 
while entropy is expressed in units of it.
The black hole's surface gravity $\kappa$ is 
replaced by the gradient of the velocity of the sound propagating upstream  
[see Fig. 1(a)], evaluated on the event horizon  \cite{Unruh81,Visser98}
\begin{eqnarray}
\kappa &=&  
 \frac d{dx} \left( v - c \right),
\label{kappa}
\end{eqnarray}  
where $v$ is the flow velocity, which is assumed positive, and where $c$ is the speed of sound 
in the comoving frame. A way to visualize (\ref{kappa}) is to think at
the rapidity of which sound wave packets diverge exponentially from the sonic horizon, 
i.e., a local Lyapunov exponent in the geometric-acoustics limit  (see Fig. 1).

\begin{figure}[b]   
\vspace{-0.1cm}
\centerline{\includegraphics[scale=0.41]{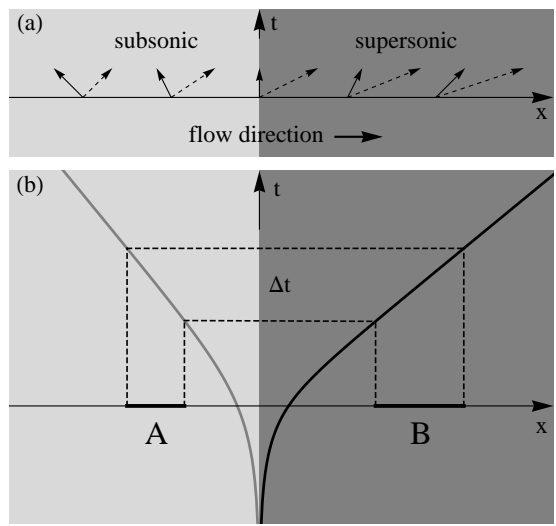}}
\vspace{-0.12cm}
\caption{(a) A fluid turning supersonic is a black hole analogue. 
Solid and dashed arrows represent local velocity vectors in a space-time diagram 
of the upstream and downstream modes, respectively.
The upstream mode has a special role in the theory as it has a bifurcation on the event 
horizon characterized by the local exponent $\kappa$. 
(b) Hawking particles are created on the event horizon in correlated pairs, which populate 
the upstream mode. The evolution of such pairs (e.g. the gray and black curves) almost 
simultaneously generated at symmetric displacements from the event horizon in the far 
past time defines a correspondence between pairs of segments, let us say $A$ and $B$, 
where  mutual information is considered.}
\label{Fig1}
\end{figure}
In this Letter, we focus on entanglement entropy production of stationary one-dimensional 
(1D) acoustic black holes. By assuming Hawking radiation (see below), the rate of which
entropy flows from the sonic horizon into subsonic and supersonic regions is given by
\begin{eqnarray}
\dot{S} =  \frac {\kappa}{12 }.
\label{EntropyProductionRate}
\end{eqnarray} 
Remarkably, this formula has a similar flavor to Pesin's formula for Kolmogorov-Sinai
entropy rate and various generalization of it, where Lyapunov exponents are 
crucial ingredients  \cite{Nicolis96,Ruelle85,Prigogine78}.
Here, however, there is no overall production of entropy, 
and mutual information between the two parts separated by the horizon forms 
 in order to compensate for the entropy produced in each part (see below).
The predictions for the entropy and mutual information 
are confirmed by an \emph{ab initio} analytical calculation in 
transonic flows of  1D ideal degenerate Fermi fluids (a microscopic model).
Finally, mutual information and entropy
could be determined in an experimental simulation of a dumb hole with 1D quantum  fluids 
 by measuring fluctuations of the number of particles.

\emph{Entropy and Mutual Information}.---Result (\ref{EntropyProductionRate}) 
follows by assuming that the collective modes propagating away 
from the horizon are thermally populated at Hawking temperature.
Each mode with wave vector  $q$ and dispersion relation $\omega(q)$ carries the entropy 
$s=x/(e^{x}-1)-\ln(1-e^{-x})$, where $x=\hbar \omega(q)/kT_{\text{bh}}$ \cite{Landau1980}. 
The entropy current (\ref{EntropyProductionRate}) is obtained by integrating the entropy 
multiplied by the group velocity, i.e., $\int_0^{\infty} s\, \frac{d\omega}{dq} \, \frac{dq}{2\pi} $.
This simplifies to $\int_0^{\infty} s\,\frac{d \omega}{2\pi}$ implying an independence on 
the actual form of the dispersion, provided this is gapless.
Instead, the entropy distributed on a large probe segment of length $L$
is calculated by assuming a local linear dispersion for the thermally populated upstream 
modes and is given by
\begin{eqnarray}
S =  \pi L k T_{\text{bh}} / 6 \hbar |v-c|.
\label{entropyvsT}
\end{eqnarray} 
In fact, Eq. (\ref{EntropyProductionRate}) has the meaning of an average entropy 
current; i.e.  $S$ 
is given by the product of (\ref{EntropyProductionRate}) and 
the time $\Delta t=L/|v-c|$ necessary for an upstream sound signal to cross it, 
provided that $\kappa \Delta t \gg 1$ in order that the thermal wavelength
associated to Hawking radiation
\begin{eqnarray}
l_{\text{H}} &=& 2 \pi |c-v| / \kappa
\label{Hawkinglength}\end{eqnarray}
could be well resolved on the length $L$.

The idea that entropy is produced in a pure quantum system may be puzzling at first 
glance.  In order to understand how this fits with a null global von Neumann entropy, 
one has to take into account correlations between the inside and outside of the dumb 
hole (see Ref. \cite{Holzhey94} for the astrophysical black hole case). 
In fact, a mutual information associated with such correlated creation events forms on 
the event horizon at twice the rate 
(\ref{EntropyProductionRate}) in order to exactly compensate for the entropy produced.

This argument can be made quantitative by recognizing that Hawking excitations are 
created in correlated pairs, which depart from the event horizon  in opposite directions 
with an average relative time uncertainty of order of $1/\kappa$ and with speed $v-c$ 
proper of the upstream sound mode. 
The evolution of such pairs indicates a correspondence between the location of a 
Hawking particle and that of his partner within an uncertainty of order of the local 
Hawking thermal length $l_{\text{H}}$.
This defines an approximate correspondence between pairs of segments, let us say $A$ 
and $B$ located on opposite sides of the event horizon [see Fig. 1(b)]. 
It is reasonable to require that the information on $A$ plus $B$ is not lost; i.e., for a 
pure quantum state the entanglement entropy \cite{Bombelli86,Holzhey94} of 
the combined system  $S_{A+B}=0$ provided $\Delta t \gg 1/\kappa$.
Moreover, since the information encoded in sound fluctuations of the upstream mode on 
$A$ and $B$ is circa the same, also the corresponding entropies, taken as measure of 
the maximal possible information content, are so. 
Introducing the mutual information as $I_{AB}=S_{A} + S_{B} - S_{A+B}$ 
helps in summarizing: if information is preserved 
\begin{eqnarray}
I_{AB} = \kappa \Delta t/6
\label{GenMutualInformation}
\end{eqnarray} 
 between corresponding segments $A$ and $B$.

\emph{Microscopic model}.---The model fluid is a degenerate 1D ideal Fermi fluid 
\cite{Giovanazzi05}. The fluid flows  in the presence of a  repulsive potential 
$V_{\text{ext}}(x)$, which acts as a 1D version of a de Laval nozzle  \cite{Giovanazzi04} 
fixing the sonic event horizon at the position  of his maximum 
$V_{\text{max}}$ (for convenience $x=0$).
The degenerate flux of incoming particles is originated from a reservoir with upper single 
particle energy $\mu_{\text{r}}$ sufficient higher than $V_{\text{max}}$ in order to have a finite
 flow over the barrier.
The potential $V_{\text{ext}}$ is assumed sufficiently smooth so that the semiclassical
description for the fluid and the consequent hydrodynamic description are appropriate 
\cite{Giovanazzi05}.
The hydrodynamic description of the sound propagation
and the associated acoustic wave equation with his gravitational analogy
are given in Ref. \cite{1DFermiHydrodynamic}. 
The potential maximum acts as a bifurcation for particles propagating 
at energy $\epsilon=V_{\text{max}}$, which is  characterized locally by the
exponent 
\begin{eqnarray}
\kappa= \sqrt{ -d^{2} (V_{\text{ext}} / m) / dx^{2}  } 
\label{kappamicro}
\end{eqnarray} 
 in a classical  description of particle propagation.
Quantum mechanics resolves such bifurcation by a finite tunneling, which
for asymptotically smooth potential barrier is always  
 consistent with a thermal smoothening of the Fermi distribution 
at temperature $k T_{\text{bh}}=\hbar \kappa /  2 \pi$.
In fact, here $k T_{\text{bh}}$ is related to top of the hill transmission and reflection 
coefficients by $|r_{\epsilon}|^{2}=1-|t_{\epsilon}|^{2}= 
\left\{ 1+\exp [ 2 \pi (\epsilon-V_{\text{max}})/\hbar\kappa] \right\}^{-1}$, which is a known 
 result \cite{Hanggi90}.
It is simple to verify that the acceleration $\kappa=d|v-c|/dx$ of the upstream sound mode, 
which propagates at the speed 
$v(x)-c(x)=\text{sign} (x) \sqrt{ 2 [ V_{\text{max}} - V_{\text{ext}}(x)] / m  } $
is fixed by the curvature of the potential hill by (\ref{kappamicro}).

For consistency with the hydrodynamic description, the energy difference $\mu_{\text{r}}-V_{\text{max}}$ 
that corresponds to twice the hydrodynamic value of $m c^2$ at the sonic horizon,
should be much bigger than $k T_{\text{bh}}$ in order that quantum corrections
remain negligible in the semiclassical approximation.
This is reminiscent of Hawking's theory of evaporating black holes where 
the back-reaction of the radiation to the gravitational metric is neglected and 
Hawking temperature is consistently assumed small compared to
the black hole mass  $M c^2$.

The thermodynamic entropy and mutual information can be extracted from the microscopic 
model by using the entanglement entropy as introduced by Bombelli et al. in \cite{Bombelli86} 
and Holzhey et al. in \cite{Holzhey94} in the context of astrophysical black holes. 
Entanglement entropy is defined relative to a certain space region as the von Neumann 
entropy $S(\rho)=- {\bf Tr} \, \rho \log \rho$, where $\rho$ is the reduced density matrix of 
a pure quantum state, made "impure" by confining it to certain space region 
\cite{Bombelli86,Holzhey94}. 
For noninteracting fermions, the entanglement entropy $S$ is directly related to the moments 
of the Full Counting Statistics \cite{Klich09}. In particular, when this last is Gaussian then $S$ 
is  just proportional to the number-fluctuation square  $\delta^2N$ 
via 
\begin{eqnarray}
S=(\pi^2/3)\;\delta^2N.
\label{SvsdN2fermi}
\end{eqnarray}
Therefore, the mutual information between the segments $A$ and $B$ is proportional to the 
correlation of the respective number fluctuations 
$I_{AB}=-(\pi^2/3)\delta N_A \delta N_B$, and the problem is reduced to the integration 
\begin{eqnarray}
\delta N_A \delta N_B &=& \int_A dx \int_B dx'  \;\; \delta \rho^{(2)}(x',x)
\label{gen_fluct_vs_2body_corr}
\end{eqnarray}
 of the two-particle correlation 
 $\delta\rho^{(2)}(x',x) = \langle \hat{\rho}(x') \hat{\rho}(x) \rangle- \rho(x')  \rho(x)$,
 which is computed below.

The many-body wave function of the fluid is the Slater determinant that corresponds 
to particles of mass $m$ coming from the left and occupying scattering wave-functions 
$\psi_k$ with wave-vector $k$ between $0$ and $k_{\text{m}}=\sqrt{2 m \mu_{\text{r}} / \hbar^2}$.
The scattering wave-functions $\psi_k$ are defined asymptotically  as 
$\psi_k(x) = \exp(i k x) + r_k  \exp(-i k x)$ for $x$ large negative and as 
$\psi_k(x) = t_k \exp(i k x) $ for $x$ large positive.
Far from the potential region, the particles move in a constant potential, which 
for the sake of simplicity is assumed zero. 
The  escape velocity from the potential  hill 
$v_{\text{esc}}=\sqrt{2 V_{\text{max}} / m}$  
corresponds to the asymptotic velocity $|c-v|$ of the upstream mode.

For a Slater determinant the two-body correlation function is related to the off-diagonal part 
of the one-body density correlation function 
$\rho^{(1)}(x,x')=\langle \Psi^{\dag}(x) \Psi(x') \rangle $ via
$\delta \rho^{(2)}(x',x) = \rho(x) \delta(x-x')-|\rho^{(1)}(x,x')|^2$, 
where $\rho(x)=\rho^{(1)}(x,x)$.
For the original zero temperature noninteracting Fermi fluid 
$\rho^{(1)}(x,x')=\int_{0}^{k_{\text{m}}} \frac {dk} {2 \pi }\; \psi^{*}_k(x) \psi_k(x')$. 
We shall start  with the computation of the fluctuations in the  subsonic region. 
We assume $A$ located far away from the event horizon in the asymptotic flat region.
Then $\rho^{(1)}(x,x')$ results in 
\begin{eqnarray}
\rho^{(1)}(x,x') &=& 
\frac{e^{i k_{\text{m}} (x'-x)}}{2\pi i (x'-x)} +\frac{e^{-i k_{\text{esc}} (x'-x)}}{2\pi i (x'-x)}
F[\pi (x'-x)/ l_{\text{H}}],\nonumber\\
\label{offdiagA}
\end{eqnarray}
where $k_{\text{esc}}=m v_{\text{esc}}/\hbar$ and $F(y)=y / \sinh(y)$ is a function that cuts 
off distances larger than $l_{\text{H}} = 2 \pi v_{\text{esc}} / \kappa$.  
For large negative coordinates $x'$ and $x$, $\rho^{(2)}$ results  in
\begin{eqnarray}
\rho^{(2)}(\Delta) = \rho_A \delta(\Delta) - \frac{ 1+F^2 -2 F \cos(2 k_{\text{F}} \Delta) }{(2\pi \Delta)^2},
\label{2inA}
\end{eqnarray}
where $\Delta=x'-x$, $k_{\text{F}}= (k_{\text{m}} + k_{\text{esc}})/2$ is the local Fermi 
wave-vector and $\rho_A=k_{\text{F}}/\pi$ is the density in $A$.
After integration of (\ref{2inA}) in $A$, the entanglement entropy results in
\begin{eqnarray}
S_{A} = \frac 1 6 \ln \left[ \frac{ l_{\text{H}} }{ \pi L_A }\sinh \left( \frac{\pi L_A}{l_{\text{H}}} \right)  \right]
+ \frac 1 3 \ln \left[ \frac{ L_A k_F }{ \pi  } \right].
\label{EntropyDistribuitedFiniteSize}
\end{eqnarray}
Note that the second term on the right side, which can be rewritten as $1/3 \ln(N_A)$ 
represents the vacuum contribution (zero temperature) to the entanglement entropy and 
corresponds to the renormalized entanglement entropy in a conformal field theory 
\cite{Holzhey94}. In the large $L_A$ limit and neglecting vacuum-type corrections the 
entropy $S_A$ is given 
\begin{eqnarray}
S_{A} = \pi  L_A / 6  l_{\text{H}}=  \kappa  \Delta t /12,
\label{EntropyDistribuited}
\end{eqnarray}
where $\Delta t = L_A/v_{\text{esc}}$. 
Thus, Eq. (\ref{EntropyDistribuited}) 
is a microscopic verification of  Eq. (\ref{entropyvsT}) and 
indirectly of the entropy production rate (\ref{EntropyProductionRate}).
The calculation of  $S_B$ gives analogue results and is not reported here.

In the following,  the mutual information $I_{AB}$ is evaluated. For this purpose, we shall 
calculate $\rho^{(2)}(x,x')$ between the opposite sides of the barrier.  
For large negative $x$ and large positive $x'$ the term in the integration for 
$\rho^{(1)}(x,x')$ proportional to $t_k$ goes to zero like $1/(x'-x)$ and is thus negligible. 
Retaining the remaining term proportional to $r^*_k t_k$ results in
$\rho^{(1)}(x,x') = e^{i k_{\text{esc}} (x'+x)} / 2 l_{\text{H}} \cosh\left[\pi (x'+x) /l_{\text{H}}\right]$. 
Thus, the two-body correlation is negative and given by
\begin{eqnarray}
\delta \rho^{(2)}(x,x')
= - 1/4 l_{\text{H}}^2 \cosh^2\left[ \pi (x'+x)/ l_{\text{H}}\right].
\label{2bodycorrelation}
\end{eqnarray}
Finally, the mutual information $I_{AB}=-(\pi^2/3)\delta N_A \delta N_B$
is obtained by the integration 
(\ref{gen_fluct_vs_2body_corr}) of (\ref{2bodycorrelation}) 
and results in 
\begin{eqnarray}
I_{AB}=\kappa \Delta t / 6
\label{MicroscopicMutualInformation}
\end{eqnarray}
as leading term for $L/l_{\text{H}} \gg 1$. 
Thus, the above {\it ab initio} calculation confirms 
the information conservation constrain for long corresponding segments.

Result (\ref{2bodycorrelation}) for the two-body correlation is very similar 
to a corresponding result for a weakly interacting Bose gas
obtained by Balbinot et al. in Ref.  \cite{Balbinot08}  [see Eq. (8) of their paper].
This correspondence further confirms the universal character of Hawking radiation.

To summarize so far, the production of entropy and mutual information 
can be extracted, in virtu of Eq. (\ref{SvsdN2fermi}),
 from the measurement of  the particle fluctuations $\delta N_A ^2$
and $\delta N_B ^2$  and of the correlations $\delta N_A \delta N_B$, respectively. 
This suggests  a method to test the entanglement entropy 
associated to Hawking radiation, that 
would also be feasible with the state of the art in measuring fluctuations.
In fact, two recent experiments \cite{Muller2010Sanner2010} 
 have demonstrated the possibility to extract the temperature $kT$ 
 from the fluctuations $\delta N ^2$  in ultracold Fermi gases
using the classical fluctuation result $\delta^2 N = N  k T / m c^2$.
In our nonequilibrium quantum dumb hole, the classical equilibrium result is replaced by
\begin{eqnarray}
\frac{\delta^2 N }{ N} =     \frac{k T_{\text{bh}}}{2 m c |c-v|} 
\label{numbersqueezingvsT}
\end{eqnarray} 
that is equivalent to Eq. (\ref{EntropyDistribuited})  by Eq. 
(\ref{SvsdN2fermi}) modulo notations.
Equation (\ref{numbersqueezingvsT}) can be also derived for 1D quantum fluids
(see supplemental material \cite{SupplementalMaterial}).

A finite reservoir temperature $kT_{\text{res}}$ calculation of (\ref{2inA}), not shown here, 
leads to
$\delta^2 N / N =  k T_{\text{bh}}\,/ 2 \,m\, c\, |c-v|+k T_{\text{res}}\,/ 2 \,m\, c\, |c+v| $. 
In deriving  this relationship, it is assumed that  $kT_{\text{res}}$ and $kT_{\text{bh}}$ 
are much smaller than $ m c^2$ at the sonic horizon. 
Moreover, the segment length should be much larger than both Hawking thermal length 
$ l_{\text{H}}$ and reservoir thermal length $l_T= \hbar (c+v) / kT_{\text{res}}$ so that 
finite size effects due to zero point fluctuations are negligible.
However, in an experimental realization with ultracold atoms, it may be necessary to 
include such contributions that correspond to
(\ref{EntropyDistribuitedFiniteSize}).

\emph{Quantum fluids}.---Mutual information may be extracted from 
experiments on 1D quantum fluids in a similar way as for 1D ideal Fermi fluids.
At thermal equilibrium and for ultralow temperatures that
the speed of sound is still close to his zero temperature value, the
entropy is related to $\delta N^2 $ by
\begin{eqnarray}
 S =   (\pi     / 3 \,\xi n )\;\delta N ^2,
\label{entropy_vs_fluctuations}
\end{eqnarray}
where $\xi=\hbar / mc$ is the vacuum correlation length and $n$ the fluid density.
Equation (\ref{entropy_vs_fluctuations}) is satisfied also when 
only one branch of the sound dispersion is thermal 
as one can see by comparing Eqs. (\ref{entropyvsT}) and (\ref{numbersqueezingvsT}).
Equation (\ref{entropy_vs_fluctuations}) generalizes (\ref{SvsdN2fermi}) and
 may have a broader range of validity.
Some progress may be gained by assuming relationship (\ref{entropy_vs_fluctuations})
for the nonequilibrium situation of our sonic black hole.
This may be the natural way to relate the 
mutual information $I_{AB}$ with the correlated fluctuations between two segments 
$A$ and $B$. I shall tentatively write
\begin{eqnarray}
S_{A+B} =   \frac{\pi} 3 \Bigl( \frac{ \delta ^2 N_A }{ \xi_A n_A}+ \frac{ \delta ^2 N_B }
{ \xi_B n_B} +\frac{ \delta N_A\delta N_B }{ \sqrt{ \xi_A n_A\xi_B n_B}}\Bigl),
\label{combined_entropy_vs_fluctuations}
\end{eqnarray}
where $\sqrt{\xi n}$ is considered a normalization factor for the fluctuations $\delta N$ 
on each homogeneous segment. 
By requiring the conservation of information $S_{A+B}=0$ 
and using Eqs. (\ref{numbersqueezingvsT}) and 
(\ref{combined_entropy_vs_fluctuations}) 
it follows that the particle number fluctuations $\delta  N_A $ and $ \delta N_B$ 
should be anticorrelated according to 
\begin{eqnarray}
\frac{\delta  N_A \delta N_B } {\sqrt {N_A N_B}}&=& -  
\frac {  k T_{\text{bh}} }{2 m \sqrt { c_A (c_A-v_A) c_B (v_B-c_B)}}.
\label{correlated_squeezing_1}
\end{eqnarray}
This expression is indeed  consistent with the prediction of
\cite{Balbinot08} and generalizes the result obtained in the microscopic model. 
In fact, using Eq. (8) of Ref. \cite{Balbinot08} for the two-particle correlation function 
$\delta\rho^{(2)}(x',x)$ 
and performing the integration
(\ref{gen_fluct_vs_2body_corr}),
it results in Eq. (\ref{correlated_squeezing_1}) as the leading term in $\Delta t$.
Thus, the above conjecture (\ref{correlated_squeezing_1}) 
together with Eq. (\ref{numbersqueezingvsT}) may be useful 
for the experimental investigation of the sonic analogue of Hawking radiation
in a 1D quantum fluid (see also the supplemental material \cite{SupplementalMaterial} 
for a possible finite reservoir temperature).

By introducing the concept of entanglement entropy production 
in transonic (1D) quantum fluids, acoustic Hawking radiation can be interpreted 
as a dissipation mechanism peculiar of a sonic horizon.

\end{document}